\documentclass[%
reprint,
 amsmath,amssymb,
 aps,
 prb,
 floatfix
]{revtex4-2}

\usepackage[english]{babel}
\usepackage{float}
\usepackage{comment}
\usepackage[dvipsnames]{xcolor}
\usepackage{graphicx}
\usepackage{dcolumn}
\usepackage{bm}
\usepackage{hyperref}
\usepackage{soul} 

\begin{document}


\title{Low-temperature thermal transport in moir\'e superlattices}

\author{Lukas P. A. Krisna}
  \email{lukas@qp.phys.sci.osaka-u.ac.jp}
\author{Takuto Kawakami}
\author{Mikito Koshino}
\affiliation{Department of Physics, Osaka University, Toyonaka, Osaka 560-0043, Japan}
\date{\today}

\begin{abstract}
We calculate the phonon thermal conductivity of various moir\'e bilayer systems using a continuum approach and the semiclassical transport theory.
When the twist angle is close to 0,
we observe a significant reduction of thermal conductivity 
in a particular low-temperature regime.
This reduction is attributed to a moir\'e-induced reconstruction of acoustic phonon bands and associated decrease of the group velocity.
Conversely, in the zero temperature limit, the thermal conductivity is enhanced by moir\'e effect, surpassing the original values in non-moir\'e counterparts.
These changes result in a characteristic temperature dependence which deviates from the quadratic behavior in intrinsic two-dimensional systems.
\end{abstract}


\maketitle

\section{Introduction}

In recent years, significant advancements have emerged in the field of two-dimensional (2D) moir\'e materials, following the discovery of various exotic phenomena in twisted bilayer systems \cite{dean2013hofstadters,hunt2013massive,cao2018correlated,cao2018unconventional,sharpe2019emergent,polshyn2019large,cao2020strange}. 
The electronic structure of these materials are strongly modulated by the emerged superlattice of the long-range periodic pattern \cite{dossantos2007graphene,li2010observation,tramblydelaissardiere2010localization,morell2010flat,luican2011single,bistritzer2011moire,moon2012energy,yankowitz2012emergence,jung2015origin}.
In a similar manner, moir\'e patterns also have notable influence on the lattice vibration.
In twisted systems, the interlayer moir\'e potential folds the original monolayer phonon bands into the superlattice Brillouin zone \cite{jiang2012acoustic, camposdelgado2013raman,cocemasov2013phonons,choi2018strong,lin2018moire,parzefall2021moire}. 
When the moir\'e superlattice period is much greater than
the atomic scale, spontaneous lattice relaxation renormalizes the phonon modes \cite{koshino2019moire,suri2021chiral,krisna2023moire,lamparski2020soliton,maity2020phonons,gadelha2021localization,quan2021phonon,maity2022chiral,lu2022low,xie2023lattice,girotto2023coupled},
and the acoustic bands are reconstructed with notable band flattening and the appearance of spectral gaps \cite{koshino2019moire,suri2021chiral,krisna2023moire}.
These changes are expected to have significant influence in the thermal transport phenomena where acoustic phonons are the dominant heat carrier.

Heat transport in 2D materials has attracted notable interest because of its superiority and its sensitivity to heterostructuring.
For example, the high thermal conductivity of suspended graphene \cite{balandin2008superior,ghosh2008extremely,cai2010thermal,chen2012thermal} is significantly decreased in the presence of substrate or other layer due to strong scattering of flexural phonons \cite{seol2010two,ghosh2010dimensional,guo2011manipulating,lindsay2011flexural,pak2016theoretical,zou2017phonon,zou2019size}.
Meanwhile, experimental measurements demonstrated further suppression in rotationally stacked graphene layers \cite{li2014thermal,han2021twist}.
Theoretical simulations suggested that 
such a thermal conductivity reduction by a twist occurs due to the enhancement of anharmonic phonon scattering or the redistribution of phonons towards higher frequencies \cite{li2018commensurate,nie2019how,mandal2022tunable,cheng2023magic,ahmed2023understanding}.
However, these studies have focused solely on room temperature and above, while the effect of moir\'e-induced phonon band reconstruction
is expected to be relevant in the low-temperature regime, and it remains unknown.

\begin{figure}
    \centering
    \includegraphics[width=\columnwidth]{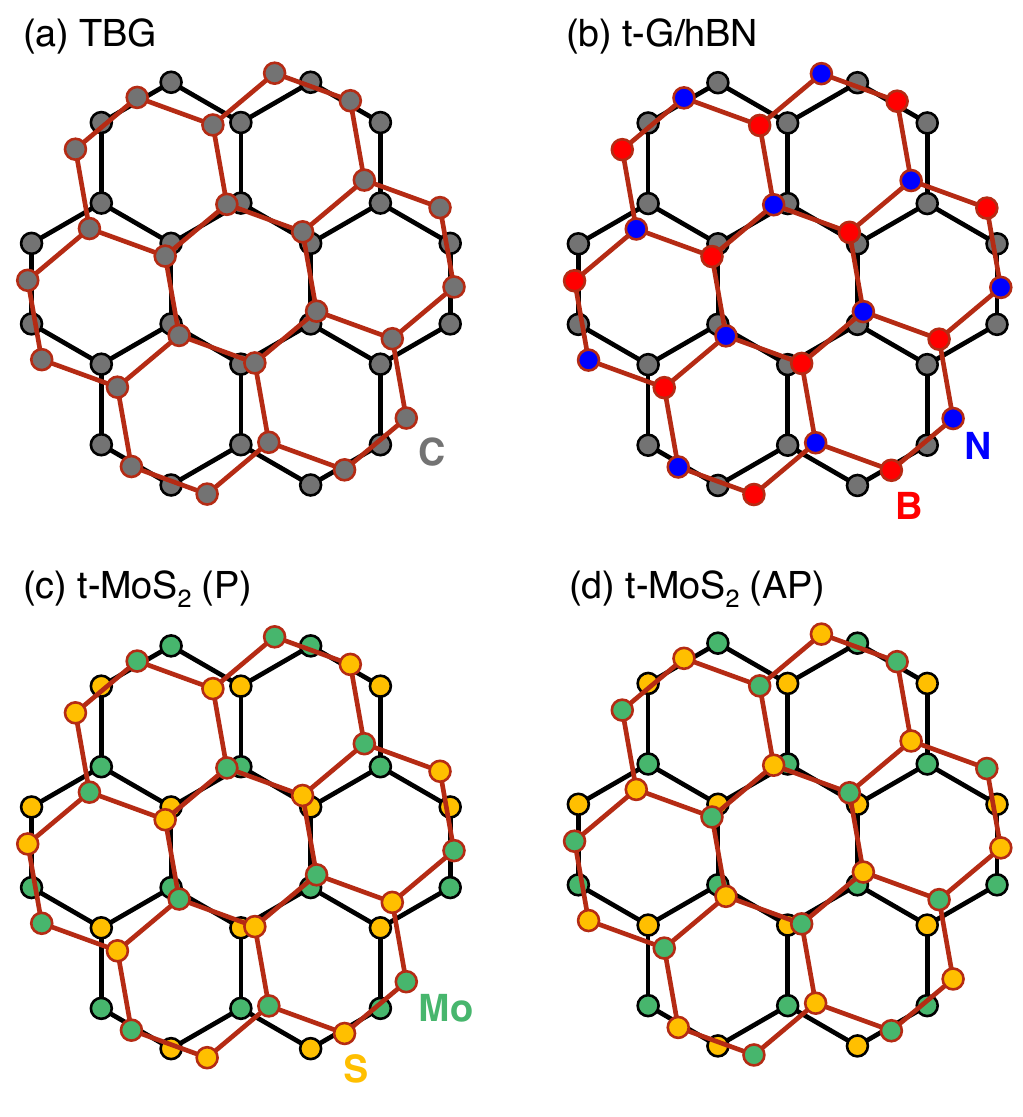}
    \caption{\label{fig:atomic}
     Top view of atomic structures of (a) twisted bilayer graphene (TBG), (b) twisted graphene/hBN (t-G/hBN), (c) parallel-stacked twisted bilayer MoS$_2$ (t-MoS$_2$ (P)), and (d) antiparallel-stacked twisted bilayer MoS$_2$ (t-MoS$_2$ (AP)).
   Each yellow circle in (c) and (d) represent overlapped double S atoms at different perpendicular positions.}
\end{figure}

In this paper, we calculate the phonon thermal conductivity of various representative twisted bilayer systems, including twisted bilayer graphene (TBG) and twisted graphene/hexagonal boron nitride (t-G/hBN), and also twisted bilayer molybdenum disulfide (t-MoS$_2$) in both parallel (P) and antiparallel (AP) stacking arrangement
[Fig.~\ref{fig:atomic}]. 
Here, we obtain the phonon band structure by 
the continuum approach, which has been extensively utilized to describe lattice relaxation and the formation of superlattice mini-bands in the phonon dispersion relation \cite{koshino2019moire,suri2021chiral,samajdar2022moire,krisna2023moire}.
Then we calculate the thermal conductivity using a semiclassical transport theory in the low temperature regime assuming a constant mean free path.
We demonstrate that the flattening of the low-energy phonons bands leads to a significant suppression of thermal conductivity up to 40\%,
resulting in a characteristic deviation from the generic quadratic temperature dependence of thermal conductivity in two-dimensional system.

This paper is organized as follows.
In Sec.~\ref{sec:method}, we introduce the continuum method to describe long-wavelength phonons, and the formulation of semiclassical transport to calculate the thermal conductivity in the low-temperature regime.
In Sec.~\ref{sec:results}, we present and discuss the calculated phonon dispersion of each system and the corresponding thermal conductivity.
Finally, a brief conclusion is given in Sec.~\ref{sec:conc}.

\section{\label{sec:method} Methods}
\subsection{Lattice geometry}

\begin{figure}
    \centering
    \includegraphics[width=\columnwidth]{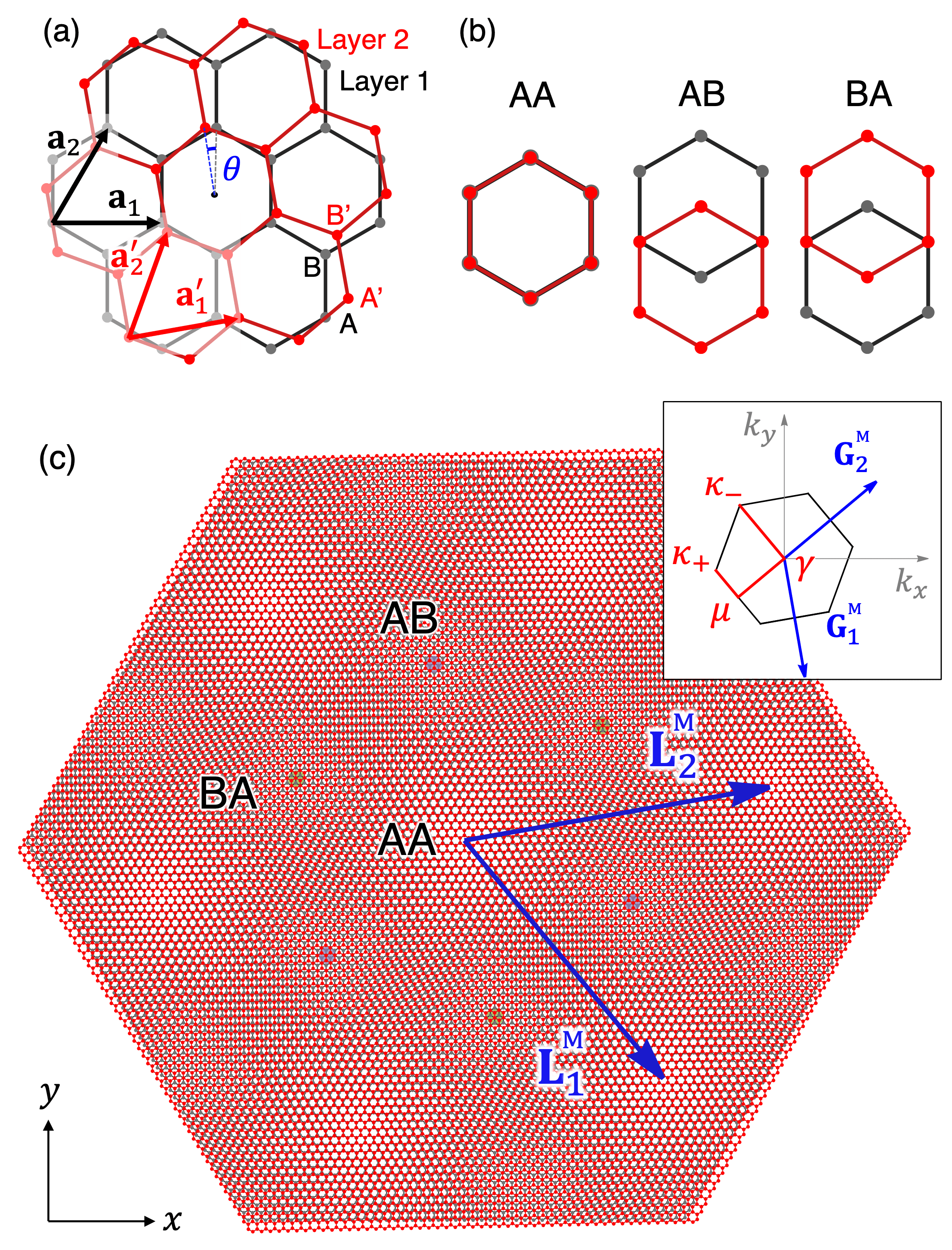}
    \caption{\label{fig:moire}
    (a) Schematic diagram of a twisted bilayer system. (b)  Three local stacking structures AA, AB, and BA. (c) Non-relaxed atomic structure of t-G/hBN with $\theta = 1.25^\circ$. The inset shows the first moir\'e Brillouin zone.}
\end{figure}

We consider a twisted bilayer system composed of two honeycomb lattice layers with generally different lattice constants, $a$ and $a'$, as illustrated in Fig.~\ref{fig:moire}(a).
The layer 2 is stacked on top of the layer 1 with relative rotation angle $\theta$ around a common honeycomb center.
We label the sublattices of layer 1 
by A and B, and that of layer 2
by A$'$ and B$'$ as in Fig.~\ref{fig:moire}(a).
The primitive lattice vectors of layer 1 are defined as $\mathbf{a}_1 = a(1,0)$ and $\mathbf{a}_2 = a(1/2,\sqrt{3}/2)$, 
and those of layer 2 are given by $\mathbf{a}'_i = \hat{M}\hat{R}\,\mathbf{a}_i\,(i=1,2)$, 
with rotation matrix $\hat{R}(\theta)$ and isotropic expansion matrix $\hat{M} = (1+\varepsilon)\hat{I}$ where $\varepsilon = (a'-a)/a$. 
The corresponding reciprocal lattice vectors for layer 1 and 2 are given by $\mathbf{b}_i$ and $\mathbf{b}'_i$, respectively, 
which satisfy $\mathbf{a}_i \cdot \mathbf{b}_j = \mathbf{a}'_i \cdot \mathbf{b}'_j = 2\pi \delta_{ij}$. 
A long-range moir\'e interference pattern appears due to a slight mismatch from a small difference in lattice constant or small twist angle.
The reciprocal lattice vectors of the pattern are given by $\mathbf{G}^{\rm M}_i = \mathbf{b}_i - \mathbf{b}'_i$, 
while the corresponding real-space lattice vectors are obtained from the condition $\mathbf{L}^{\rm M}_i \cdot \mathbf{G}^{\rm M}_j = 2\pi\delta_{ij}$.
The moir\'e period $L_{\rm M} = |\mathbf{L}^{\rm M}_i|$ can be expressed as
\begin{equation}
    L_{\rm M} = a \frac{1 + \varepsilon}{\sqrt{\varepsilon^2 + 2(1 + \varepsilon)(1 - \cos{\theta})}}.
\end{equation}

For honeycomb lattice components, we take graphene, hexagonal boron nitride, and molybdenum disulfide with lattice constant of $a \approx$ 0.246 nm, 0.2504 nm, and 0.317 nm, respectively.
In this paper, we consider 
twisted bilayer graphene (TBG) 
and twisted bilayer molybdenum disulfide (t-MoS$_2$) as examples of homo-bilayer ($\varepsilon=0$)
and also twisted graphene/hexagonal boron nitride (t-G/hBN) as a heterobilayer ($\varepsilon\neq0$).
In Fig.~\ref{fig:moire}(c), we illustrate the formation of moir\'e superlattice in t-G/hBN with $\theta = 1.25^\circ$ where a lattice constant difference $\varepsilon\approx 1.8\%$ and the twist angle produce a moir\'e pattern of $L_{\rm M} = |\mathbf{L}^{\rm M}_i| \approx 8.8$ nm.

Across the moir\'e pattern, the local stacking structure changes smoothly at the atomic scale.
At a given position $\mathbf{r}$, it is characterized by the phase difference 

$\bm{\varphi}(\mathbf{r})=(\varphi_1(\mathbf{r}),\varphi_2(\mathbf{r}))$, where $\varphi_j(\mathbf{r})$ is defined as 
\begin{equation}\label{eq:phase}
   \varphi_j (\mathbf{r}) = (\mathbf{b}_j - \mathbf{b}'_j) \cdot \mathbf{r} = \mathbf{G}^\text{M}_j\cdot \mathbf{r}.
\end{equation}
For example, $\bm{\varphi} = (0, 0)$, $(2\pi/3, 2\pi/3)$ and $(4\pi/3, 4\pi/3)$ correspond to
AA (complete alignment of the honeycomb lattices), 
AB (B$'$-site of layer 2 on top of A-site of layer 1), and 
BA (A$'$-site of layer 2 on top of B-site of layer 1), respectively [Fig.~\ref{fig:moire}(b)].

We note that, when stacking two monolayers with different sublattice atoms, e.g., t-MoS$_2$, there are two possible stacking configuration, parallel (P) and antiparallel (AP), 
which are related by 180$^\circ$ rotation of layer 1.
In the parallel stacking, the two layers have identical atoms at A and A$'$ (B and B$'$),
whereas in the antiparallel stacking, the two layers have different atoms on the same sublattice sites.


\subsection{Continuum model}

We describe the in-plane lattice vibration
in the twisted bilayer systems using a continuum approach \cite{koshino2019moire,suri2021chiral,krisna2023moire}.
In this framework, we consider a smoothly-varying displacement field
$\mathbf{u}^{(l)}(\mathbf{r},t)$ at layer $l (=1,2)$, which represents the shift of the atom at $\mathbf{r}$ in the original rigid lattice, and express the Lagrangian as a functional of $\mathbf{u}^{(l)}(\mathbf{r},t)$.
Let us consider the in-plane atomic displacement $\mathbf{u}^{(l)}= (u_x^{(l)},u_y^{(l)})$.
In the presence of the lattice distortion, the phase difference [Eq.~\eqref{eq:phase}] is modified as
\begin{equation}\label{eq:phasediff}
    \varphi_j(\mathbf{r},t) = \mathbf{G}^\text{M}_j\cdot\left[\mathbf{r}-\mathbf{u}^+(\mathbf{r},t)/2\right]
    +\bar{\mathbf{b}}_j\cdot\mathbf{u}^-(\mathbf{r},t),
\end{equation} 
where we have defined
the (interlayer) symmetric and antisymmetric displacements $\mathbf{u}^\pm = \mathbf{u}^{(2)}\pm\mathbf{u}^{(1)}$,
and $\bar{\mathbf{b}}_j=(\mathbf{b}_j+\mathbf{b}'_j)/2$.

The Lagrangian is explicitly written as $L = T - (U_E + U_B)$, where $T$ is the kinetic energy, $U_E$ is the elastic energy, and $U_B$ is the interlayer binding energy.
The total interlayer binding energy is calculated as the integral of the local interlayer binding energy over the whole system,
\begin{equation}
 U_B =  \int V[\bm{\varphi}(\mathbf{r},t)]\,d^2\mathbf{r},
\end{equation}
where $V[\bm{\varphi}]$ is the interlayer binding energy per area of the non-rotated bilayer with an interlayer registry specified by $\bm{\varphi}$ \cite{nam2017lattice,krisna2023moire}.
By 120$^\circ$ symmetry of the system, $V[\bm{\varphi}]$ should be written within the lowest harmonics as
\begin{equation}
    V[\bm{\varphi}(\mathbf{r},t)] = \sum_{j=1}^3 2V_0 \cos\left[\varphi_j\left(\mathbf{r},t\right) +\varphi_0\right],
\end{equation}
where
$\varphi_3 = -\varphi_1 - \varphi_2$,
$\varphi_0$ is a material-dependent phase, and the constant energy offset is neglected.
The energy of AA, AB, and BA bilayer stacking is then given by
$V_{AA} =6 V_0 \cos(\varphi_0)$, 
$V_{AB}= 6 V_0 \cos(\varphi_0 + \tfrac{2}{3}\pi)$, and 
$V_{BA}=6 V_0 \cos(\varphi_0 - \tfrac{2}{3}\pi)$.
Here we obtain the parameters $(V_0,\varphi_0)$ from the relative values of $V_{AA}$, $V_{AB}$, and $V_{BA}$ found in the literature.
Table \ref{tab:binding-par} lists the values of $(V_0,\varphi_0)$ and the corresponding $(V_{AA},V_{AB},V_{BA})$ for the considered systems in this paper.

\begin{table*}[]
    \caption{\label{tab:binding-par}
    Parameters for interlayer binding energy, $(V_0,\varphi_0)$, and the corresponding $(V_{AA},V_{AB},V_{BA})$ for the considered systems in this paper.
    Note that the origin of interlayer binding energy is arbitrary.
    }
    \begin{ruledtabular}
    \begin{tabular}{lddddd}
               & \multicolumn{1}{c}{$V_0$ (eV/nm$^2$)} 
               & \multicolumn{1}{c}{$\varphi_0$} 
               & \multicolumn{1}{c}{$V_{AA}$ (eV/nm$^2$)} 
               & \multicolumn{1}{c}{$V_{AB}$ (eV/nm$^2$)}  
               & \multicolumn{1}{c}{$V_{BA}$ (eV/nm$^2$)}  \\
        \hline
         TBG\footnotemark[1]  &  0.160  &  0  & 0.961 & -0.481 & -0.481 \\
         t-G/hBN\footnotemark[2] & 0.202 & 0.956 & 0.700 & -1.208 & 0.509 \\
         t-MoS$_2$ (P)\footnotemark[3] & 0.0889 & 0 & 0.533 & -0.267 & -0.267 \\
         t-MoS$_2$ (AP)\footnotemark[3] & -0.0801 & -0.805 & -0.333 & -0.133 & 0.467
    \end{tabular}
    \end{ruledtabular}
    \footnotetext[1]{Ref.~\cite{popov2011commensurate,lebedeva2011interlayer}}
    \footnotetext[2]{Ref.~\cite{sanjose2014spontaneous}}
    \footnotetext[3]{Ref.~\cite{enaldiev2020stacking,weston2020atomic}}
\end{table*}

The energy cost for intralayer distortion, $U_E$ is given by standard expression of elastic theory \cite{suzuura2002phonons,nam2017lattice},
\begin{align}\label{eq:2-uel}
    U_E = \sum_{l=1}^2 \frac{1}{2}\int
    &(\lambda^{(l)}+\mu^{(l)})\left(u^{(l)}_{xx} + u^{(l)}_{yy}\right)^2 \nonumber \\
    &+\mu^{(l)}\left[\left(u^{(l)}_{xx} - u^{(l)}_{yy}\right)^2+4\left(u^{(l)}_{xy}\right)^2\right]d^2\mathbf{r},
\end{align}
where 
$u_{ij}^{(l)} = (\partial_i u_j^{(l)} + \partial_j u_i^{(l)})/2$ is the strain tensor,
and $\lambda^{(l)}$ and $\mu^{(l)}$ are the Lam\'e parameters for layer $l$ which are given in Table \ref{tab:intralayer_parameters}.
\begin{table}[]
    \caption{\label{tab:intralayer_parameters}
    Lam\'e parameter and mass density used in the calculation.}
    \begin{ruledtabular}
    \begin{tabular}{lddd}
               & \multicolumn{1}{c}{$\lambda$ (eV/\AA$^2$)} 
               & \multicolumn{1}{c}{$\mu$ (eV/\AA$^2$)} 
               & \multicolumn{1}{c}{$\rho$ ($10^{-8}$ g/cm$^2$)} \\
        \hline
         Graphene\footnotemark[1]  & 3.25 & 9.57 & 7.61 \\
         hBN\footnotemark[2] & 3.5 & 7.8 & 7.59 \\
         MoS$_2$\footnotemark[3] & 4.23 & 4.23 & 30.5 \\
    \end{tabular}
    \end{ruledtabular}
    \footnotetext[1]{Ref.~\cite{zakharchenko2009finite,jung2015origin}}
    \footnotetext[2]{Ref.~\cite{sachs2011adhesion,jung2015origin}}
    \footnotetext[3]{Ref.~\cite{enaldiev2020stacking,weston2020atomic}}
\end{table}
The kinetic energy is expressed as
\begin{equation}\label{eq:2-kinetic}
   T = \sum_{l=1}^2\int\frac{1}{2}\rho^{(l)}\left(\dot{u}_x^{(l)2}+ \dot{u}_y^{(l)2}\right)d^2\mathbf{r},
\end{equation}
where $\rho^{(l)}$ is the mass density of layer $l$.
The mass density for graphene, hBN, and MoS$_2$ are given in Table \ref{tab:intralayer_parameters}.
For convenience, we define averaged materials parameters
\begin{align}\label{eq:lambda_mu_rho}
 &   \lambda = \frac{1}{2}(\lambda^{(1)}+\lambda^{(2)}), 
    \quad 
    \mu = \frac{1}{2}(\mu^{(1)}+\mu^{(2)}),
     \nonumber\\
    & \rho = \frac{1}{2}(\rho^{(1)}+\rho^{(2)}).
\end{align}


The equation of motion is given by the  Euler-Lagrange equation for the Lagrangian $L$.
It can be effectively decoupled in terms of the symmetric and the antisymmetric components $\mathbf{u}^\pm$
\cite{koshino2019moire,krisna2023moire}.
The hybridization between $\mathbf{u}^\pm$ is caused by the difference of the parameters $\rho^{(l)}, \lambda^{(l)}, \mu^{(l)}$ between two layers,
and therefore it is present only in heterobilayers \cite{krisna2023moire}.
In bilayers composed of materials with closely matched physical parameters, such as t-G/hBN, these effects are minor and will be neglected in the following argument.

The effect of the moir\'e interlayer coupling is much more significant in the antisymmetric part $\mathbf{u}^-$
than in the symmetric part 
 $\mathbf{u}^+$,
as the phase difference $\varphi_i$ in Eq.~\eqref{eq:phasediff} is more sensitive to $\mathbf{u}^-$ because of 
$|\bar{\mathbf{b}_j}|\gg|\mathbf{G}^{\rm M}_j|$ \cite{krisna2023moire}.
By neglecting $\mathbf{u}^+$ in $\varphi_i$, the equation of motion for $\mathbf{u}^+$ becomes equivalent to that for the single-layer honeycomb lattice with the averaged parameters of Eq.~\eqref{eq:lambda_mu_rho}.
 As a result, the symmetric phonon modes are simply represented by the longitudinal (LA) and transverse (TA) acoustic modes with phonon velocity 
\begin{equation}\label{eq:vlvt}
    v_{\rm L} = \sqrt{\frac{\lambda+2\mu}{\rho}}\quad\text{and}\quad 
    v_{\rm T} = \sqrt{\frac{\mu}{\rho}},
\end{equation}
respectively.

\begin{figure}
    \centering
    \includegraphics[width=\columnwidth]{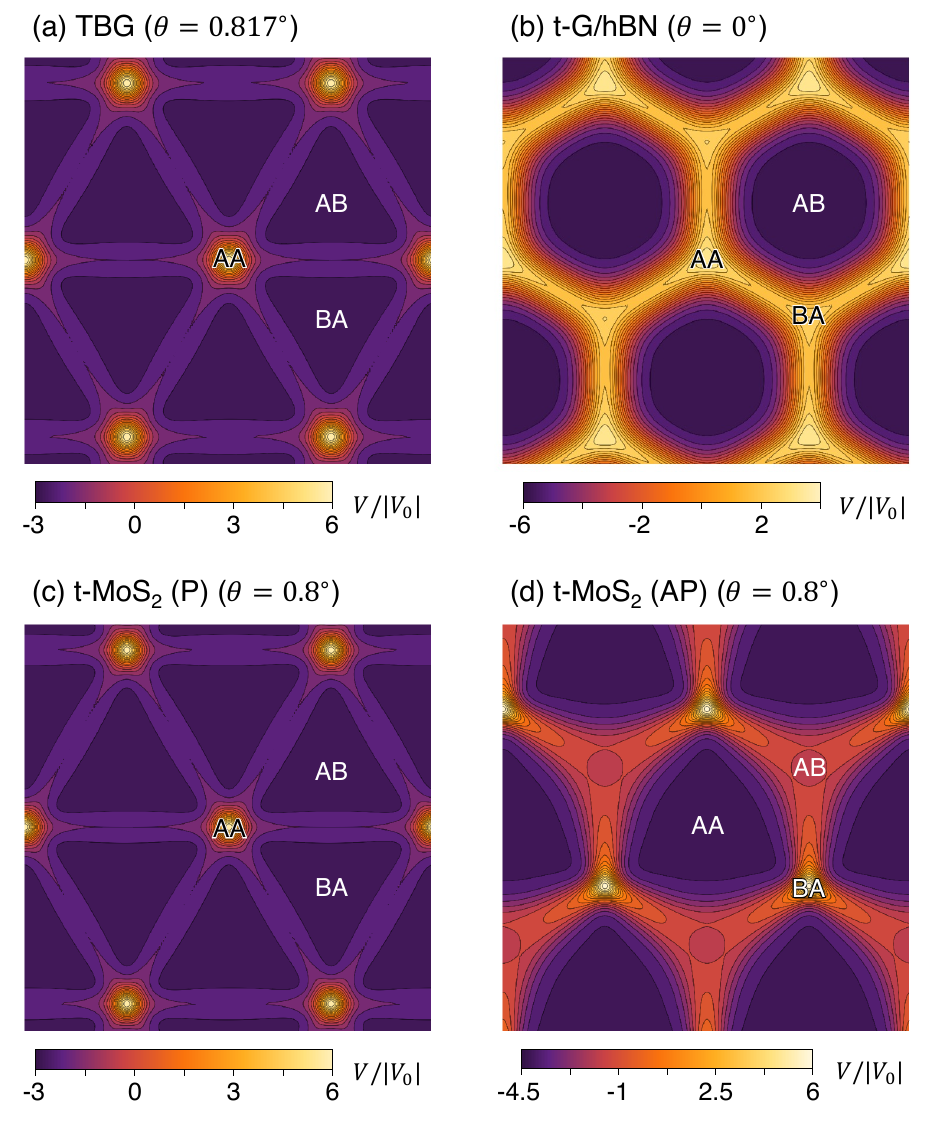}
    \caption{
    Contour plot of the interlayer binding energy $V(\mathbf{r})$ in the relaxed structure of (a) 0.817$^\circ$ TBG, (b) 0$^\circ$ t-G/hBN, (c) 0.8$^\circ$ t-MoS$_2$ (P), and (d) 0.8$^\circ$ t-MoS$_2$ (AP).
    }
    \label{fig:domain_relax}
\end{figure}

For the antisymmetric part $\mathbf{u}^-$,
we consider the solution for  in the form of
\begin{equation}\label{eq:upertb}
    \mathbf{u}^-(\mathbf{r},t) = \mathbf{u}^-_0(\mathbf{r}) + \delta\mathbf{u}^-(\mathbf{r},t),
\end{equation}
where 
$\mathbf{u}^-_0(\mathbf{r})$ gives the relaxed state to minimize $U_B + U_E$ and $\delta\mathbf{u}^-(\mathbf{r},t)$ is a small vibration around $\mathbf{u}^-_0$. 
The relaxed displacement field $\mathbf{u}^-_0(\mathbf{r})$ is obtained by self-consistent iteration \cite{nam2017lattice}. 
Figure \ref{fig:domain_relax} presents contour plots of the interlayer binding energy in the relaxed state for the considered moir\'e systems. 
Each system exhibits a characteristic domain pattern, which is determined by relative stabilities among AA, AB and BA stacking configurations (Table \ref{tab:binding-par}).
In TBG [Fig.~\ref{fig:domain_relax}(a)] and parallel t-MoS$_2$ [Fig.~\ref{fig:domain_relax}(c)], the relaxed structure reveals a triangular pattern comprising AB and BA stacking regions, indicative of the energetic equivalence between these configurations
\cite{nam2017lattice,carr2018relaxation}.
In t-G/hBN [Fig.~\ref{fig:domain_relax}(b)], on the other hand, only the AB stacking dominates in the relaxed structure as it is the only structure with the lowest energy,
resulting in a honeycomb domain pattern
\cite{sanjose2014spontaneous,woods2014commensurate}.
The antiparallel t-MoS$_2$ [Fig.~\ref{fig:domain_relax}(d)] exhibits a honeycomb structure reminiscent of t-G/hBN, albeit with a distortion of the hexagonal domains into triangles. This distortion arises from the relatively small energy difference between the most stable AA stacking and the second stable AB stacking (see, Table \ref{tab:binding-par}), resulting in a significant broadening of domain walls around the AB position
\cite{enaldiev2020stacking,weston2020atomic}.

The vibration modes around the relaxed structure can be obtained as follows.
We define the Fourier transform of 
$\mathbf{u}^-_0(\mathbf{r})$ and $\delta\mathbf{u}^-(\mathbf{r},t)$ as,
\begin{align}
\mathbf{u}^-_0(\mathbf{r}) &= 
   \sum_{\mathbf{G}} \mathbf{u}^-_{0,\mathbf{G}} e^{i\mathbf{G}\cdot\mathbf{r}}, \\
    \delta\mathbf{u}^-(\mathbf{r},t) &=
   \frac{1}{\sqrt{S}}
    \sum_\mathbf{G}\sum_\mathbf{q}
    \delta\mathbf{u}_{\mathbf{q}+\mathbf{G}}^-(t)
    e^{i(\mathbf{q}+\mathbf{G})\cdot\mathbf{r}},
\end{align}
where $\mathbf{G} = m\mathbf{G}^\text{M}_1+n\mathbf{G}^\text{M}_2$ are the moir\'{e} reciprocal lattice vectors, $\mathbf{q}$ is the wave vector within MBZ, and $S$ is the system's total area as normalization factor. The equation of motion in the momentum-space is then given as
\begin{equation}\label{eq:eom}
\frac{\rho}{2}
\frac{d^2}{dt^2}\delta\mathbf{u}_{\mathbf{q}+\mathbf{G}}^- =- \sum_{\mathbf{G}'}\hat{D}_\mathbf{q}(\mathbf{G},\mathbf{G}')\delta\mathbf{u}^-_{\mathbf{q}+\mathbf{G}'},
\end{equation}
where
\begin{equation}
    \hat{D}_\mathbf{q}(\mathbf{G},\mathbf{G}')=(1/2)\hat{K}_{\mathbf{q}+\mathbf{G}}\delta_{\mathbf{G},\mathbf{G}'}+\hat{V}_{\mathbf{G}'-\mathbf{G}}
\end{equation}
is the dynamical matrix.
The matrix $\hat{K}$ is defined as
\begin{align}
    \hat{K}_\mathbf{q} =
    \begin{pmatrix}
    (\lambda + 2\mu) q_x^2 +\mu q_y^2 & (\lambda + \mu) q_x q_y \\
   (\lambda + \mu) q_x q_y & (\lambda + 2\mu) q_y^2 + \mu q_x^2
    \end{pmatrix},
\end{align}
and the matrix $\hat{V}$ is defined as
\begin{equation}\label{eq:V-matrix}
    \hat{V}_\mathbf{G} = 
    -2V_0 
    \sum_{j=1}^3
    h_\mathbf{G}^j
    \begin{pmatrix}
    \bar{b}_{j,x}\bar{b}_{j,x} & \bar{b}_{j,x}\bar{b}_{j,y} \\
    \bar{b}_{j,y}\bar{b}_{j,x} & \bar{b}_{j,y}\bar{b}_{j,y}
    \end{pmatrix},
\end{equation}
where $h^j_\mathbf{G}$ are the Fourier components of
\begin{equation}\label{eq_cos_FT}
    \cos\left[\mathbf{G}^\text{M}_j\cdot\mathbf{r} + 
    \bar{\mathbf{b}}_j\cdot\mathbf{u}^{-}_0(\mathbf{r}) + \varphi_0 \right] = \sum_\mathbf{G}
    h_\mathbf{G}^j e^{i\mathbf{G}\cdot\mathbf{r}}.
\end{equation}
Equation \eqref{eq:eom} is solved by diagonalizing the dynamical matrix $\hat{D}_\mathbf{q}$ to obtain the phonon frequency $\omega_{n,\mathbf{q}}$ of mode $n$ at $\mathbf{q}$. 
In solving this, we take sufficient cut-off in the reciprocal space, so that the result converges.

\subsection{Phonon thermal conductivity}

Based on linearized Boltzmann transport theory with relaxation time approximation \cite{ziman2001electrons}, the phonon thermal conductivity of 2D isotropic material can be expressed as
\begin{equation}
    \label{eq:kappa}
    \kappa = \frac{1}{S}\sum_{n,\textbf{q}}
    \frac{1}{2}v^2_{n,\textbf{q}}\tau_{n,\textbf{q}}\hbar\omega_{n,\textbf{q}}\frac{\partial f_0(\omega_{n,\textbf{q}})}{\partial T},
\end{equation}
where $S$ is the total area of the system, 
$f_0(\omega) = 1/[\exp(\hbar\omega/k_B T) - 1]$ is the Bose-Einstein distribution function, 
$v_{n,\mathbf{q}}=|\bm{\nabla}_\mathbf{q}\omega_{n,\mathbf{q}}|$ is the phonon velocity, $\tau_{n,\mathbf{q}}$ is the relaxation time, and the summation is taken over all of the mode index $n$ and wave vector $\mathbf{q}$.
Here, we note that the isotropic behavior is a consequence of the three-fold rotation symmetry and the thickness of the 2D system had been factored out from the expression of $\kappa$.

\begin{figure}
    \centering
    \includegraphics[width=0.95\columnwidth]{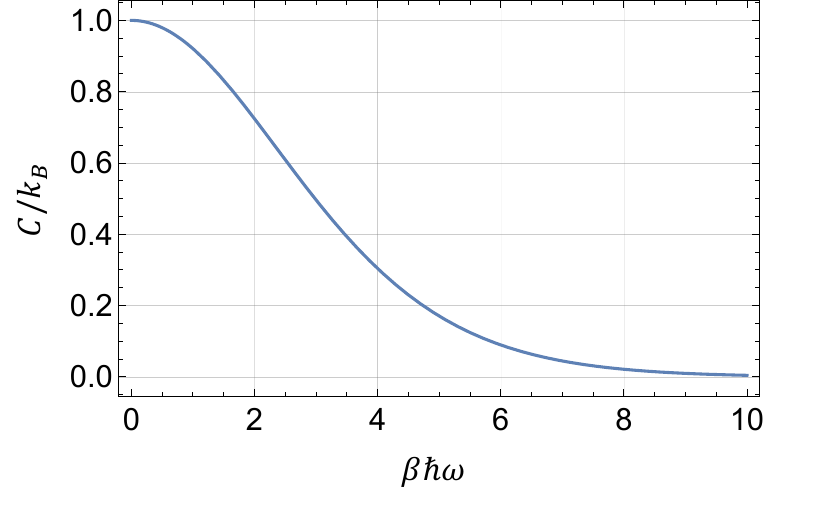}
    \caption{Plot of the spectral heat capacity function $C(\omega,T)$ [Eq.~\eqref{eq:C}] as a function of $\beta\hbar\omega$ where $\beta=1/(k_B T)$.}
    \label{fig:cplot}
\end{figure}

The relaxation time, $\tau_{n,\mathbf{q}}$, describes various scattering mechanisms that limit the mean free path of the phonon which is defined as $\Lambda_{n,\mathbf{q}} \equiv \tau_{n,\mathbf{q}} v_{n,\mathbf{q}}$.
At low temperature, scattering due to geometric boundary is the dominant scattering mechanism and the mean free path no longer depends on the phonons frequency and wavelength, i.e., $\Lambda_{n,\mathbf{q}} = \Lambda$ (constant) \cite{ziman2001electrons}.
For example, in graphitic systems, $\Lambda$ is determined from the size and shape of the sample or the grain boundaries, and it well describes the thermal conductivity for up to 100 K \cite{prasher2008thermal,heremans1985thermal,klemens1994thermal}.
In moir\'e systems, the superlattice period is generally observed to be varying across a single sample \cite{uri2020mapping,benschop2021measuring} and this would disrupt the propagation of phonon modes \cite{ochoa2022degradation}.
In such a case, $\Lambda$ can be regarded as a typical length where the moir\'e pattern remains uniform.
Henceforth, we treat $\Lambda$ as a phenomenological constant to be determined from direct measurements.
The thermal conductivity is then entirely governed by the harmonic properties, and it can be written as
\begin{equation}\label{eq:kappa-int}
    \kappa = \frac{\Lambda}{2}\int_0^\infty \tilde{n}(\omega) C(\omega, T) d\omega,
\end{equation}
where velocity-weighted density of states (VDOS) $\tilde{n}$ is defined as
\begin{equation}\label{eq:vdist}
    \tilde{n}(\omega)=\frac{1}{S}\sum_{n,\mathbf{q}}\delta(\omega-\omega_{n,\mathbf{q}})v_{n,\mathbf{q}}
\end{equation}
and spectral heat capacity $C$ is defined as
\begin{align}\label{eq:C}
    C(\omega, T) &= \hbar\omega \frac{\partial f_0(\omega)}{\partial T}     
    = k_B \left[\frac{\beta\hbar \omega/2}{\sinh(\beta\hbar \omega/2)} \right]^2,
\end{align}
where $\beta = 1/(k_B T)$.
In Eq.~\eqref{eq:kappa-int}, $\tilde{n}(\omega)$ contains all information regarding the phonon dispersion while $C(\omega,T)$ acts as a weight function. 
The function $C(\omega,T)$ is plotted in Fig.~\ref{fig:cplot}; it equals to $k_B$ when $\omega=0$, and decays exponentially for $\hbar\omega \gtrsim 2k_B T$.

In general, the thermal conductivity is contributed by all phonon modes including the out-of-plane (flexural) phonons as well as in-plane modes. 
In this work, we will treat in-plane and out-of-plane phonon modes independently, 
since the coupling between these only appears as higher-order terms of the equation of motion \cite{xie2023lattice}, 
leading to only minor effects on the overall phonon spectrum. 
Also, the effect of the moiré pattern on out-of-plane phonons is expected to be smaller than that on in-plane phonons, 
since interlayer in-plane displacement directly modifies the local atomic registry in the moiré pattern, while the out-of-plane displacement does not [23]. 
Thus, we will focus on the effect of in-plane phonons on the thermal conductivity, assuming that the contribution of out-of-plane phonons remains at the intrinsic value for a single layer. 
It is also known that in multilayered systems or in the presence of substrate, the out-of-plane modes are strongly suppressed \cite{guo2011manipulating,lindsay2011flexural,pak2016theoretical,zou2017phonon,zou2019size} especially at low temperature \cite{seol2010two}.

Since only the in-plane antisymmetric phonon modes are strongly affected by the interlayer moir\'e potential, it is useful to write $\kappa$ and $\tilde{n}$ in terms of the symmetric and antisymmetric components
\begin{align}
    \kappa = \kappa^+ + \kappa^-, \quad
    \tilde{n} = \tilde{n}^+ + \tilde{n}^-,
\end{align}
which takes the contribution from the symmetric ($+$) and antisymmetric ($-$) phonon modes separately.
In the absence of the interlayer moir\'e coupling
(i.e., two independent monolayers),
they are given as
\begin{align}\label{eq:non-coupling-xi}
    \tilde{n}^+_{\rm NC} &= \tilde{n}^-_{\rm NC} = \tfrac{1}{2}\tilde{n}_{\rm NC} 
    = \frac{\omega}{2\pi {\bar v}}, \\
    \label{eq:non-coupling-kappa}
    \kappa^+_{\rm NC} &= \kappa^-_{\rm NC} = \tfrac{1}{2}\kappa_{\rm NC} 
     = \frac{3 \Lambda \zeta(3)}{2 \pi \hbar^2 {\bar v}}k_B^3 T^2,
\end{align}
where ${\bar v}^{-1} = v_{\rm L}^{-1} + v_{\rm T}^{-1}$, $\zeta(n)$ is the Riemann zeta function, and
NC stands for `non-coupled'.
The quadratic temperature dependence of the thermal conductivity is a characteristic of linear acoustic phonon-dominated thermal transport of 2D systems in the low-temperature regime \cite{seol2010two}.

\section{\label{sec:results} Results and discussion}
\subsection{TBG and t-G/hBN}

\begin{figure*}
    \centering
    \includegraphics[width=\textwidth]{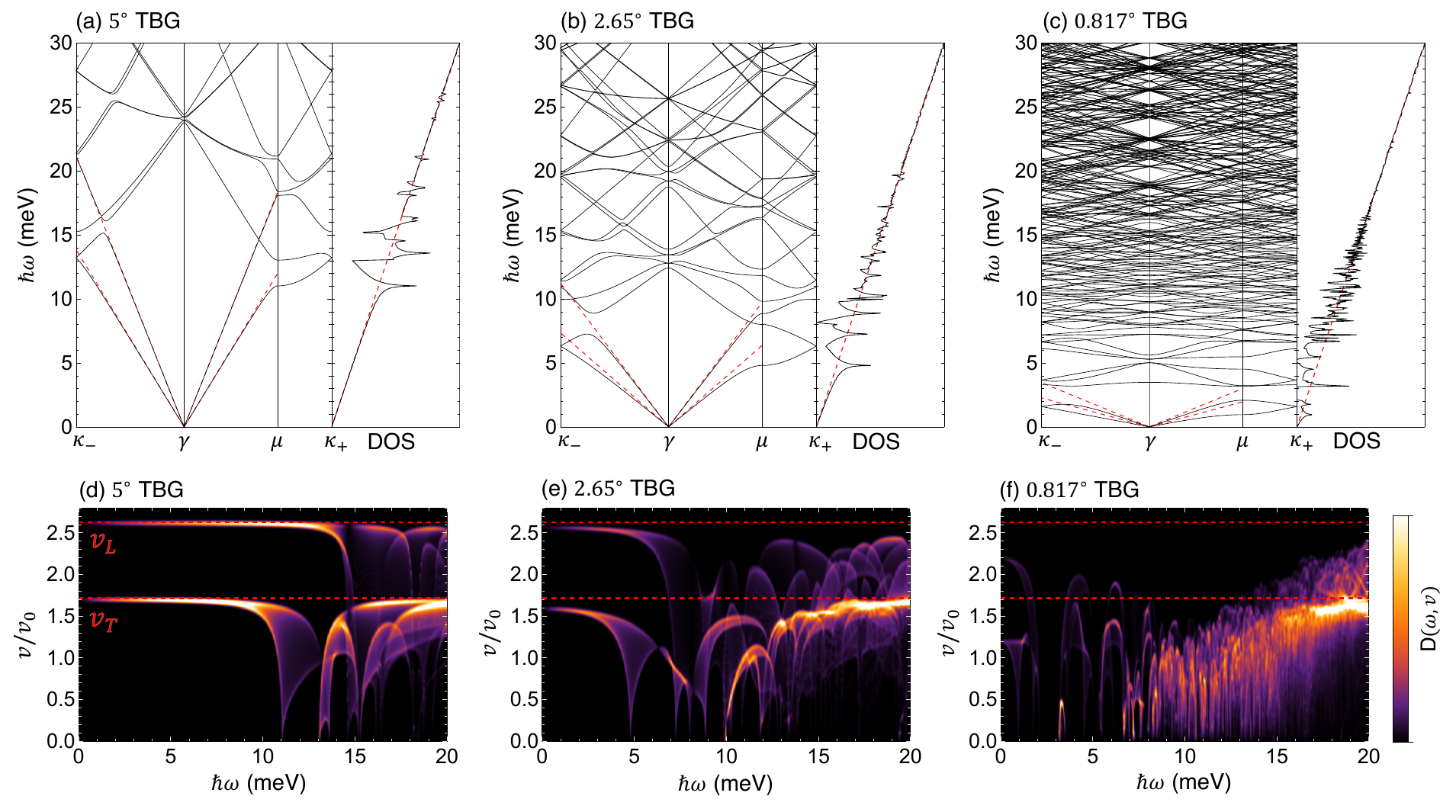}
    \caption{
    Band structure of the interlayer antisymmetric phonon modes (black line) and the symmetric modes (red-dashed line) for (a) 5$^\circ$, (b) 2.65$^\circ$, (c) 0.817$^\circ$ TBG. 
    The right panel plots the density of states. (d)-(f) Corresponding 
    plots for $D(\omega,v)$, the
    density of states on the frequency-velocity space [see, Eq.~\eqref{eq:density_2d}].
    }
    \label{fig:ph_tbg}
\end{figure*}
\begin{figure*}
    \centering
    \includegraphics[width=\textwidth]{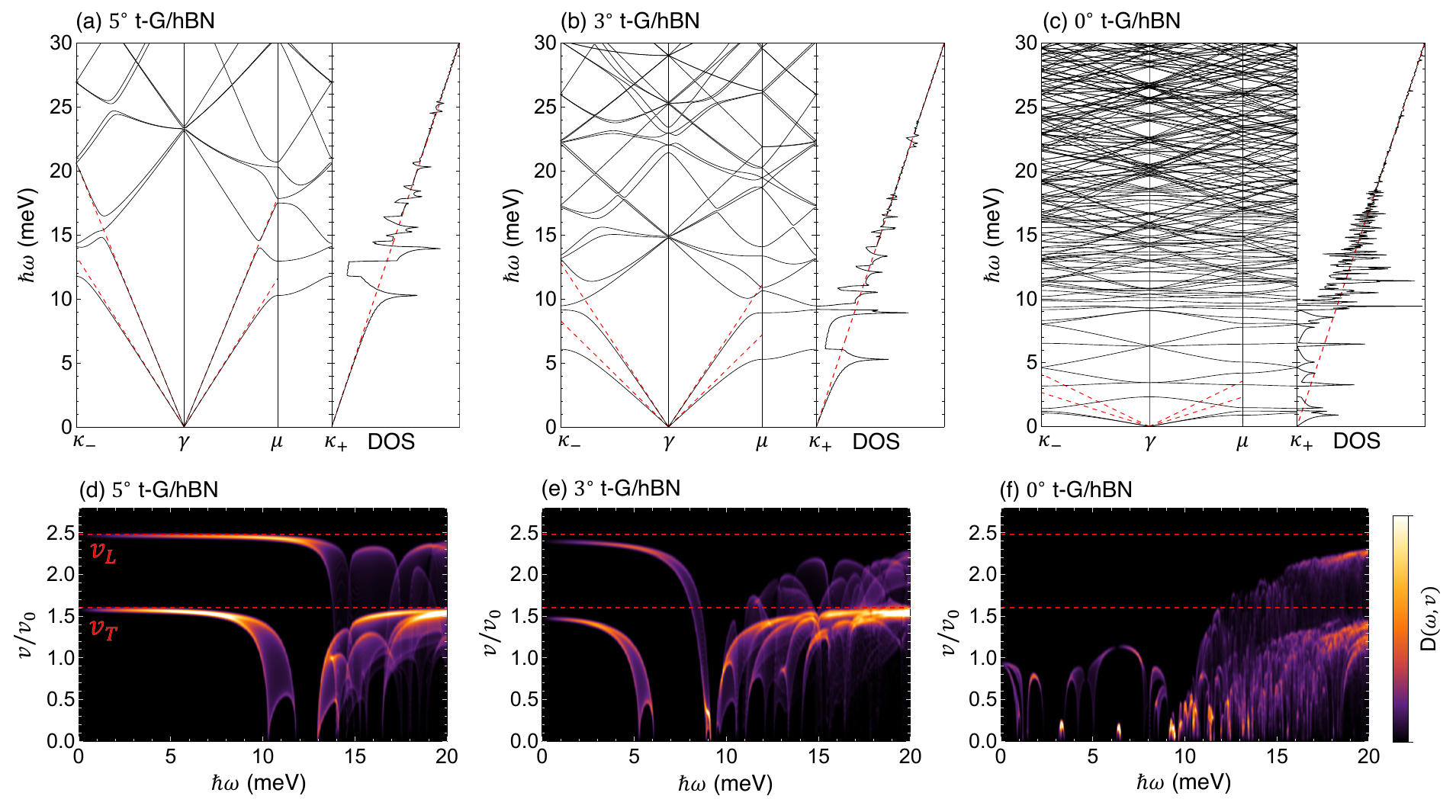}
    \caption{
    Plots of the band structure
    and $D(\omega,v)$
    similar to Fig.~\ref{fig:ph_tbg}, for t-G/hBN with $\theta = 5^\circ$,  3$^\circ$, 0$^\circ$. 
    }
    \label{fig:ph_ghbn}
\end{figure*}

Figure \ref{fig:ph_tbg} shows the calculated
band dispersion and density of states for
antisymmetric phonon modes in TBG with $\theta = $ 5$^\circ$, 2.65$^\circ$ and 0.817$^\circ$.  Figure \ref{fig:ph_ghbn} presents similar plots for t-G/hBN with $\theta = $ 5$^\circ$, 3$^\circ$, and 0$^\circ$.
The phonon dispersion is plotted along the high-symmetry line in the moir\'e Brillouin zone (MBZ) as illustrated in Fig.~\ref{fig:moire}(c).
In each panel, the red-dashed line indicate the 
dispersion (only the lowest two branches shown) and the density of states in the non-coupled case.
Here we can see that the interlayer coupling significantly modifies the band structure of the antisymmetric modes in the energy range below $\sim$20 meV.
This reconstruction is characterized by sharp peaks in the density of states due to the band flattening, which becomes more significant at smaller twist angle.
At small angle, in particular, mini band gaps also emerge at the few-meV scale. 
These gaps are attributed to the frequency modulations of string-like excitations that dominate the low-frequency modes of moir\'e superlattice systems \cite{krisna2023moire}.
Meanwhile, since the moir\'e interlayer potential does not couple the symmetric modes, the corresponding band structure is equivalent to those of the non-coupled case (red-dashed lines).

To provide a more comprehensive understanding of the band flattening effects, we present two-dimensional density maps in Fig.~\ref{fig:ph_tbg}(d)-(f) and Fig.~\ref{fig:ph_ghbn}(d)-(f), which plot
the distribution of antisymmetric phonon modes on a space of frequency and velocity, or
\begin{equation}\label{eq:density_2d}
    D(\omega,v) = \sum_{n,\mathbf{q}} \delta(\omega-\omega_{n,\mathbf{q}})
    \delta(v-v_{n,\mathbf{q}}).
\end{equation}

Here, the Dirac delta function is approximated by a Lorentzian function where the full-width at half-maximum is between 0.02 and 0.05 meV depending on the twist angle.

In each panel, the vertical axis is scaled by a factor of 
\begin{equation}\label{eq:v0}
    v_0 = \sqrt{\lambda/\rho}
\end{equation} 
and the color brightness is a linear scale of
$D(\omega,v)$.
The horizontal red-dashed lines correspond to the velocity of the TA ($v_{\rm T}$) and LA ($v_{\rm L}$) phonons of a single layer [Eq.~\eqref{eq:vlvt}].
The flattening of phonon bands can be immediately seen as a distribution of phonon modes below $v_{\rm T}$ and $v_{\rm L}$.
In 0.817$^\circ$ TBG [Fig.~\ref{fig:ph_tbg}(f)]
and 0$^\circ$ t-G/hBN [Fig.~\ref{fig:ph_ghbn}(f)],
particularly, the signals in $\hbar\omega < 15$ meV
predominantly falls below the line of $v_{\rm T}$, i.e., nearly all of the low-energy antisymmetric phonons become slower than the original acoustic phonons in its non-moir\'e counterpart.
Note that the distribution of symmetric phonon states (not shown) sticks to the $v_L$ and $v_T$ lines, where the intensity increases linearly with energy.

\begin{figure}
    \centering
    \includegraphics[width=0.9\columnwidth]{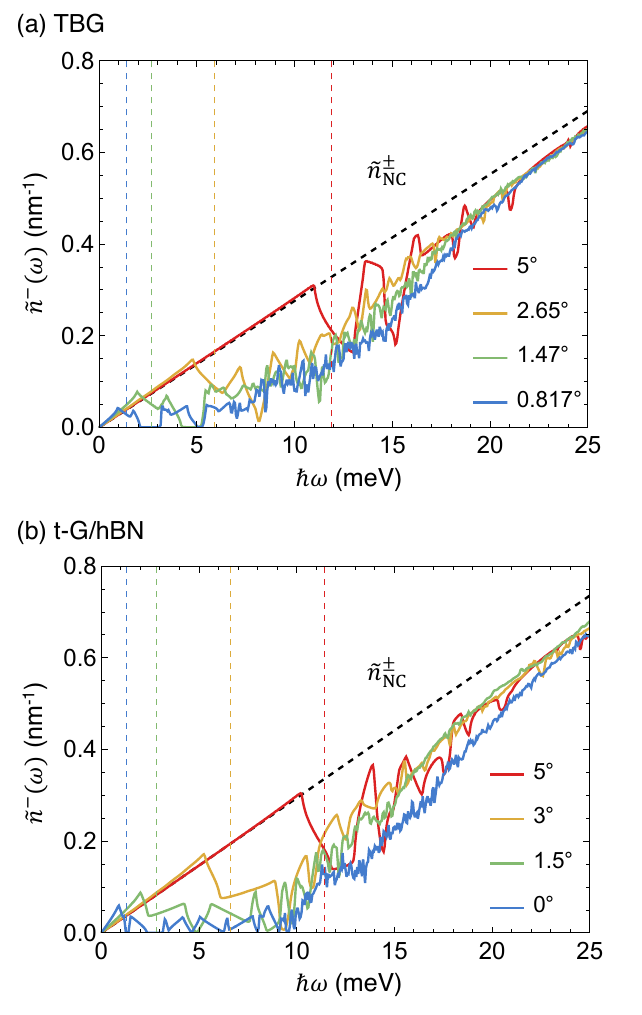}
    \caption{
(a) Velocity-weighted density of states (VDOS) 
of the antisymmetric modes, $\tilde{n}^-(\omega)$, 
in TBGs (colored lines). Each color represent different twist angles. 
Colored vertical lines correspond to $\hbar\omega_{\rm edge} \approx 2\pi\hbar v_1/(\sqrt{3}L_{\rm M})$ for each twist angle.
Black dashed lines represents $\tilde{n}^+_{\rm NC}=\tilde{n}^-_{\rm NC}$ 
for a non-coupled bilayer [Eq.~\eqref{eq:non-coupling-xi}].
The VDOS for symmetric phonons $\tilde{n}^+$ is equal to $\tilde{n}^\pm_{\rm NC}$.
(b) Similar plots for t-G/hBN.
}
    \label{fig:vdos}
\end{figure}

Figure \ref{fig:vdos}(a) and (b) show $\tilde{n}^-(\omega)$ (the VDOS of the antisymmetric phonon modes)
of TBG and t-G/hBN, respectively, with various small twist angles.
The effect of the moir\'e coupling is observed as a difference from a black-dashed line, which represents $\tilde{n}^+_{\rm NC}=\tilde{n}^-_{\rm NC} \propto \omega$ [Eq.~\eqref{eq:non-coupling-xi}].
While VDOS is proportional to both phonon density of states and velocity, we find that the reduction of phonon velocity is more significant than the sharpening of density of states,
leading to a suppression of $\tilde{n}^-$ over a wide range of phonon frequency. 
We also find that the linear behavior of $\tilde{n}^-$  remains in the low frequency region, which corresponds to the linear dispersion part of the lowest moir\'e phonon band in 
$\omega < \omega_{\rm edge} \approx 2\pi v_1/(\sqrt{3}L_{\rm M})$ where $v_1$ is the phonon velocity for the lowest band near the MBZ center.
For the symmetric phonon modes (not shown),
we have $\tilde{n}^+(\omega) = \tilde{n}^\pm_{\rm NC} (\omega)$, because they are not influenced by the moir\'e coupling.

\begin{figure*}
    \centering
    \includegraphics[width=0.8\textwidth]{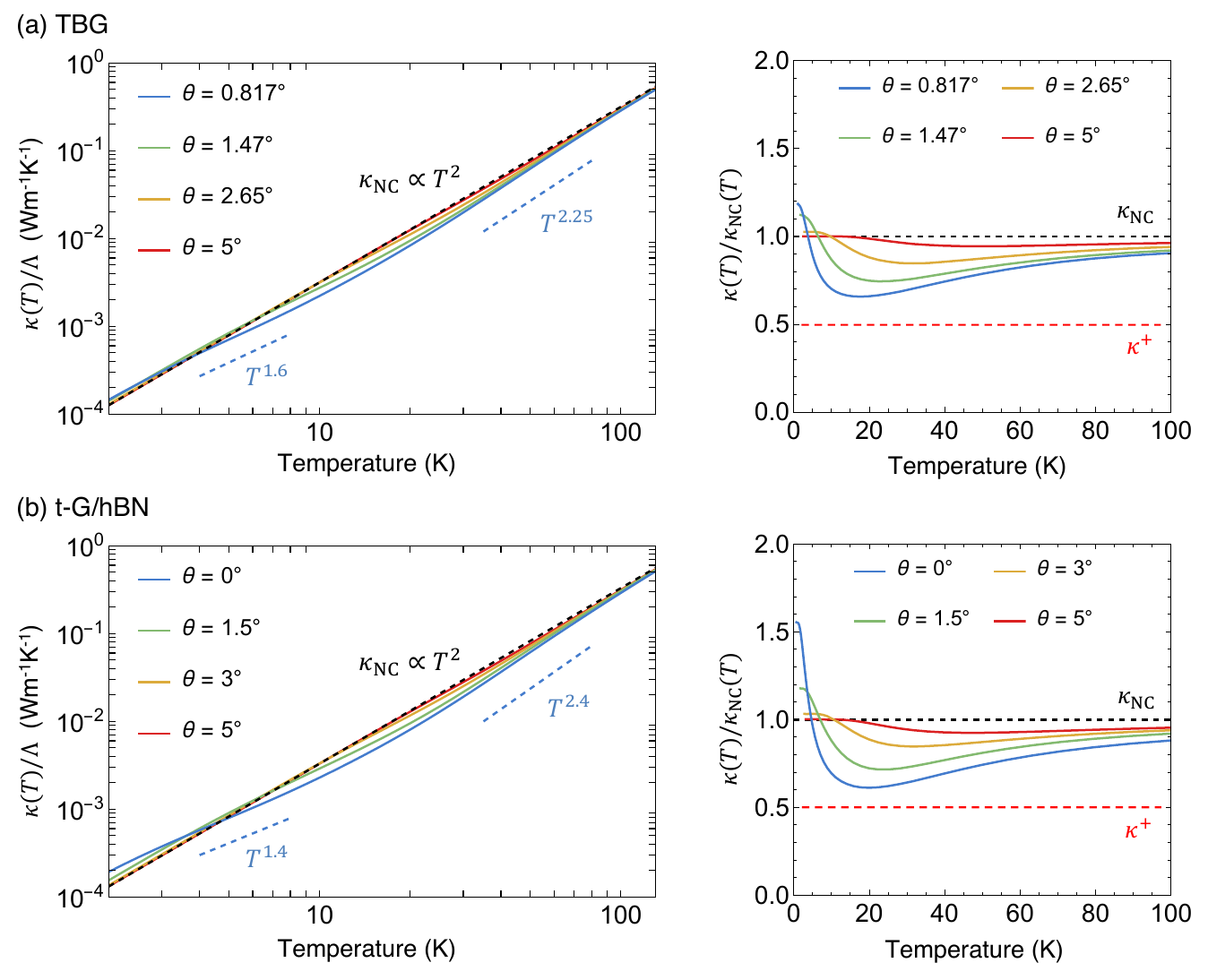}
    \caption{\label{fig:kappa}
    (a) Thermal conductivity in TBGs with various twist angles.
    The left panel shows the thermal conductivity $\kappa(T)$ scaled by the mean free path $\Lambda$.
    The right panel shows the relative thermal conductivity to the non-coupled bilayer case, $\kappa_{\rm NC}(T)$ [Eq.~\eqref{eq:non-coupling-kappa}]. (b) Similar plots for t-G/hBN.
    }
\end{figure*}

Figure \ref{fig:kappa}(a) and (b) summarise the calculated thermal 
conductivity $\kappa(T)(= \kappa^+ + \kappa^-)$ of TBG and t-G/hBN, respectively.
In each figure, the left panel shows a log-log plot of $\kappa(T)$ 
for $2 < T < 110$ K.
Here the vertical axis is scaled by the constant mean-free-path length $\Lambda$, which is assumed to be independent of temperature.
The colored lines correspond to different twist angles. The black-dashed line represents the non-coupled bilayer case, $\kappa_{\rm NC}$, which is proportional to $T^2$ as given in Eq.~\eqref{eq:non-coupling-kappa}.
Here, we find that thermal conductivity is suppressed from the non-coupled case, notably around 20 K,
while it converges towards the non-coupled value as increasing temperatures.

To better understand the change from the intrinsic graphene, we plot the relative thermal conductivity, $\kappa(T)/\kappa_{\rm NC}(T)$, in the right panel of Fig.~\ref{fig:kappa}(a) and (b).
Here, suppression of thermal conductivity is represented by a value below unity.
We can see that the suppression occurs over a wide temperature range except near the low-temperature limit.
This suppression becomes more pronounced as the twist angle is smaller, where the largest reduction of up to $\sim$35\% for 0.817$^\circ$ TBG and up to $\sim$40\% for 0$^\circ$ t-G/hBN takes place at around 20 K.
This can be understood since at around 20 K, heat transport is carried by phonons with energy below $\sim$20 meV ($\beta\hbar\omega \approx 6$) [see Fig.~\ref{fig:cplot}].
Within this energy range, the contribution of the asymmetric modes are almost completely absent [Fig.~\ref{fig:vdos}] and the thermal conductivity is almost entirely came from the symmetric phonon modes, $\kappa^+ = \kappa_{\rm NC}/2$ (red-dashed line).

In the zero-temperature limit, we observe that the relative thermal conductivity $\kappa/\kappa_{\rm NC}$ rapidly rises and even exceeds 1, indicating that the thermal transport is {\it enhanced} by the moir\'e effect.
The reason for this phenomenon is explained as follows.
Within this temperature range, only phonons in the two lowest moir\'e phonon bands in $\omega < \omega_{\rm edge}$ become relevant in the thermal transport equation.
These phonon modes have a linear dispersion,
where the VDOS is given by $\omega/(2\pi v)$
with the corresponding group velocity $v$ [Eq.~\eqref{eq:non-coupling-xi} for the noncoupled case].
Since the velocities of the antisymmetric phonon modes are significantly reduced by the moir\'e effects [see Fig.~\ref{fig:ph_tbg} and Fig.~\ref{fig:ph_ghbn}], the inverse relation leads to an enhancement of the VDOS, and hence of the thermal conductivity.
In decreasing the twist angle,
these phonon velocities are monotonically decreased, and eventually converges towards a finite value in the small angle limit 
\cite{koshino2019moire,gao2022symmetry}. This sets the upper bound of the relative thermal conductivity in the $T \to 0$ limit.

We also find that the overall modifications of the thermal conductivity by the moir\'e effect results in a change of the power coefficient $\alpha$ in $\kappa(T) \propto T^\alpha$, as seen from the logarithmic plot in the left panel of Fig.~\ref{fig:kappa}(a) and (b).
In the absence of moir\'e interlayer coupling, thermal conductivity has quadratic temperature dependence ($\alpha = 2$), which comes from the linear dispersion of the original acoustic phonons.
However, the enhancement in the low $T$ limit and the subsequent suppression in higher $T$ decrease
the power coefficient from 2.
For example, it is given by $\alpha \approx 1.6$ in 0.817$^\circ$ TBG and $\alpha \approx 1.4$ in 0$^\circ$ t-G/hBN within the temperature range of 4 K to 8 K.
As temperature increases further, moir\'e effect starts to fade out and thermal conductivity returns towards the original value. This requires the power coefficient to be larger than 2.
For example, it is given by $\alpha \approx 2.25$ for 0.817$^\circ$ TBG and $\alpha \approx 2.4$ for 0$^\circ$ for t-G/hBN which occurs in the range from 35 K to 80 K.

\subsection{t-MoS$_2$}

\begin{figure*}
    \centering
    \includegraphics[width=0.8\textwidth]{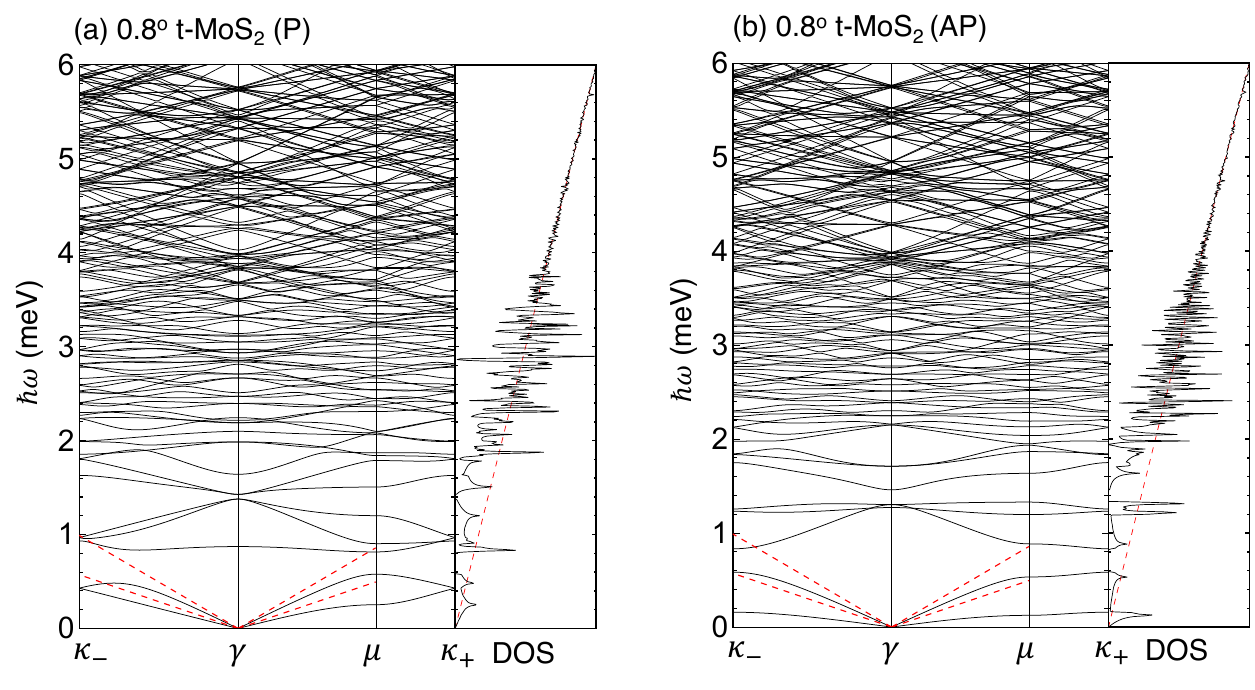}
    \caption{Phonon dispersion and density of states of the antisymmetric modes (black line) and the symmetric modes (red-dashed line) for (a) parallel-stacked (P) and (b) antiparallel-stacked (AP) 0.8$^\circ$ t-MoS$_2$.}
    \label{fig:phonon_tmdc}
\end{figure*}

The calculated phonon dispersion for t-MoS$_2$ is shown in Fig.~\ref{fig:phonon_tmdc} for both parallel and antiparallel stacking case with twist angle 0.8$^\circ$.
We find that the antisymmetric phonon modes of the parallel-stacked case closely resembles that of TBG [Fig.~\ref{fig:ph_tbg}(c)].
This similarity occurs because 
TBG and t-MoS$_2$ share triangular domain wall structures [see Fig.~\ref{fig:domain_relax}(a) and (c)].
In an analogous way, the phonon band structure of the antiparallel t-MoS$_2$ 
resembles those of t-G/hBN, reflecting a common honeycomb domain-wall structure [Fig.~\ref{fig:domain_relax}(b) and (d)]. 
This is a natural result because the moir\'e phonon band structure is qualitatively reproduced by an effective mass-spring model for domain-wall motion \cite{krisna2023moire}, and hence it is primarily
determined by the geometrical structure of domain walls (triangular or honeycomb).
We also observe that the characteristic energy scale of the moir\'e phonon bands in t-MoS$_2$ is much smaller than in TBG and t-G/hBN, because the original acoustic phonon velocity in MoS$_2$ is much lower than that of graphene and hBN.

\begin{figure}
    \centering
    \includegraphics[width=0.85\columnwidth]{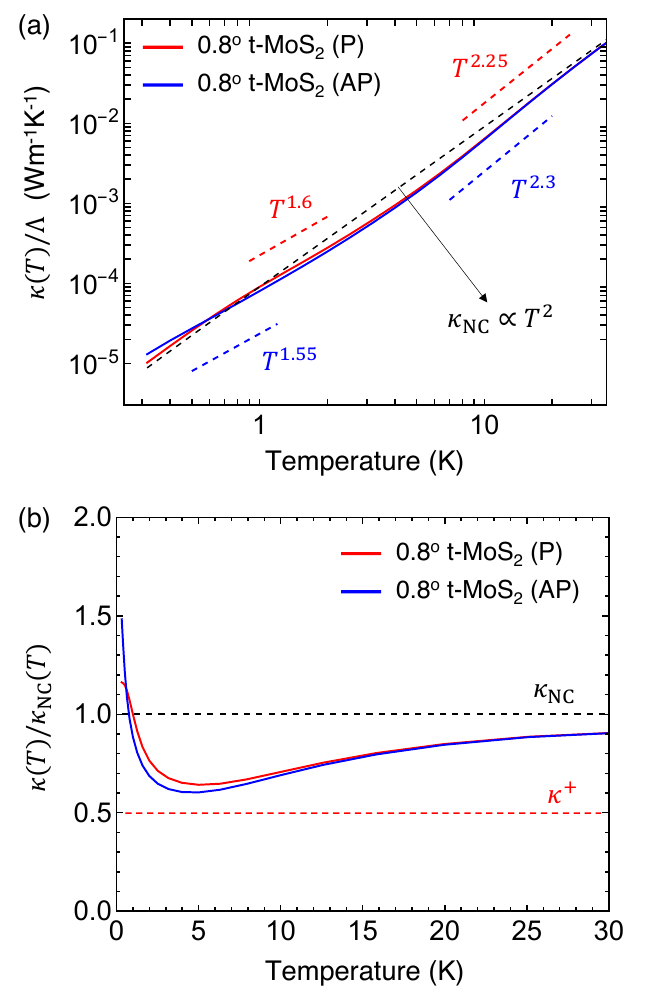}
    \caption{(a) Thermal conductivity of t-MoS$_2$ with $\theta=0.8^\circ$ for the parallel case (red line) and antiparallel case (blue line). (b) Relative thermal conductivity to the noncoupling case, $\kappa_{\rm NC}(T)$.}
    \label{fig:kappa_tmdc}
\end{figure}

In Fig.~\ref{fig:kappa_tmdc}, we plot the calculated thermal conductivity of 
t-MoS$_2$ (P and AP) in a parallel manner to the TBG and t-G/hBN case.
Here we find an overall reduction of thermal conductivity and the enhancement near $T \to 0$ just as in TBG and t-G/hBN.
However, the characteristic temperature range is much lower than  TBG and t-G/hBN because of the smaller energy scale of the corresponding moir\'e phonons.
At $T \sim$ 4 K, the thermal conductivity is reduced up to around 35\% and 40\% in the P and AP
cases, respectively.
In the limit of $T \to 0$, we find that the AP case exhibits greater enhancement of $\kappa$ than in the P case.
This is attributed to the smaller phonon velocity $v$ in the lowest branch in the AP case [Fig.~\ref{fig:phonon_tmdc}]
and the fact that $\kappa$ for the linear band regime is proportional to $1/v$ as argued in the previous section.
Accordingly, we have the corresponding change of the power coefficient $(\alpha)$ in $\kappa(T) \propto T^\alpha$
in a similar manner to TBG and t-G/hBN.

\section{\label{sec:conc} Conclusion}
We have studied the phonon thermal conductivity of various moir\'e superlattice systems in the low-temperature regime.

Using the continuum model, we showed that half of the in-plane phonon modes are hybridized by the interlayer moir\'e potential with notable reduction in their group velocity, particularly at small twist angles.
This leads to the decrease of thermal conductivity across a wide temperature range.
For example, in the case of 0$^\circ$ t-G/hBN, the highest reduction occurs at around 20 K, for up to 40\%, and this reduction slowly subsides at higher temperatures until it returns to the original intrinsic value at around 100 K.
On the contrary, at very low temperatures, thermal conductivity is shown to be inversely related to the phonon velocity, and therefore it becomes enhanced.
This continuous change over temperature leads to a power-law deviation from the original quadratic temperature dependence, that is, the exponent $\alpha$ is less than 2 from 4 K to 20 K and greater than 2 from 20 K to 80 K in the case of 0$^\circ$ t-G/hBN.
This behavior also holds in other moir\'e systems, though the temperature ranges in which it appears depend on the original acoustic phonon velocity of the constituent monolayers.

We expect future observation of the characteristic power-law deviation to be strong evidence of moir\'e effect on low-energy phonons, further highlighting its importance when investigating phonon-related phenomena in moir\'e systems.
Thermal conductivity measurements at low temperatures down to a few kelvins have been demonstrated, for example, in carbon nanotubes \cite{kim2001thermal} and thin graphite \cite{machida2020phonon}.
While distinguishing the contribution of in-plane and out-of-plane modes in thermal transport experiments is generally challenging, one could consider cases where the mean free path of the in-plane modes is significantly larger, such as in a $\mu$m-wide sample supported or sandwiched by substrates \cite{seol2010two,bae2013ballistic,pak2016theoretical,zou2017phonon,zou2019size}.

Lastly, the inclusion of anharmonic scattering between moir\'e phonons and the presence of moir\'e-scale disorder \cite{ochoa2022degradation,nakatsuji2022moire} is also expected to further decrease the thermal conductivity.
We leave this investigation for future study.

\begin{acknowledgments}
This work was supported by JST SPRING, Grant Number JPMJSP2138, JSPS KAKENHI Grants No.~JP21H05236, and No.~JP21H05232,  No.~JP20H01840, and by JST CREST Grant No.~JPMJCR20T3, Japan.
\end{acknowledgments}

\bibliography{bibliography}

\end{document}